# SEARCH FOR ALIGNED EVENTS IN MUON GROUPS DEEP UNDERGROUND

A.L.Tsyabuk, R.A.Mukhamedshin and Yu.V.Stenkin
*Institute for Nuclear Research of Russian Academy of Sciences*
*60th October Anniv., Prosp.,7a ,Moscow 117312, RUSSIA*
*Presenter: Yu.V.Stenkin(stenkin@sci.lebedev.ru),* rus-stenkin-Y-abs3-he12-poster

A search for aligned events has been done throughout the muon group events measured with Baksan Underground Scintillation Telescope (BUST) during a long period of time. Only groups of multiplicity >3 for muon threshold energies equal to 0.25; 0.85; 2.0 and 3.2 TeV were selected for the analysis. A distribution of the events on the alignment parameter λ has been obtained and compared with the results of Monte-Carlo simulation made for this experiment. Upper limit for aligned muon groups flux is given.

## 1. Introduction

The muon bundle data cover an energy range between the data of direct methods (balloons, satellites) and the Extensive Air Showers (EAS) data: from $\sim 10^{13}$ and up to $\sim 10^{17}$ eV for primary cosmic ray (i.e. includes the "knee"). High-energy muons are produced in the highest energy part of the EAS cascade at high altitudes. The muon bundle events give information on the transverse momenta of the secondaries in high-energy collisions, which affects a lateral spread of muons in the bundles. The lateral structure of multiple events underground also affects very much the observed muon multiplicity spectrum. It is worthwhile to use this information for an estimation of the chemical composition of primary cosmic rays. On the other hand, the "knee" energy range is related to interesting phenomena observed in cosmic rays, in particular, to the phenomenon of the alignment of most energetic subcores of gamma-ray-hadron (γ-h) families (particles of highest energies in the central EAS core) found in the "Pamir" emulsion chamber experiment [1,2] related to a coplanar particle production at $E_0 > 10^{16}$ eV. In this work we continued our search [3] for aligned events among the muon groups observed in Baksan Underground Scintillation Telescope (BUST) for a set of energy thresholds. The first search for aligned muon bundle events has been done in MACRO detector [4] without certain results.

## 2. The experiment

The BUST detector [5] is located in Baksan Valley (North Caucuses, Russia) at 1700 m. a. s. l. altitude in the underground laboratory at a distance of 550 *m* from the tunnel entrance. Its effective depth is 850 $hg/cm^2$ and the effective muon threshold energy is 220 GeV. The depth varies from h~800 $hg/cm^2$ for near vertical directions up to h~6000 $hg/cm^2$ for slant trajectories where the energy threshold is about 6 - 10 TeV. The telescope looks like a four-storey building with a size of 16.7×16.7×11.1 $m^3$. Four vertical scintillator layers and four horizontal scintillator planes are formed by liquid scintillation detectors of standard type. The total number of detectors is 3150. The standard detector (0.7×0.7×0.3 $m^3$) consists of aluminium tank filled with a liquid scintillator viewed by a 6-inch diameter PMT (FEU-49).

In this experiment additional requirements were applied to select the so-called *aligned* events. We used the standard λ parameter introduced by the "Pamir" experiment [1] and varying from ~1 for aligned events to negative values for isotropic events. Inclined tracks with zenith angles inside a range of 50-70° were selected to make muon energy thresholds higher. In this work we studied 4 groups of events with a different angles **θ** and time **t** as shown in Table 1.

**Table 1**. Data selection parameters.

| $E_{th}$, TeV | 0.25 | 0.85 | 2.0 | 3.2 |
|---|---|---|---|---|
| h, hg/cm$^2$ | 850 | 2500 | 4000 | 5000 |
| $\theta°$ | 0÷70 | 50÷70 | 50÷70 | 50÷70 |
| t, years | 0.79 | 7.67 | 7.67 | 7.67 |

## 3. Results

The experimental normalised distributions on the λ parameter are shown in Fig.1(a,b and c) for visible muon multiplicity *m*=4, 5 and 6 correspondingly. Each picture shows the λ distributions for $E_{th}$=0.25; 0.85; 2.0 and 3,2 TeV. There are also shown distributions on λ of the simulated absolutely random spots distributed homogeneously on the BUST area. As one can see we have no any significant access near λ=1. The simulated distributions reveal a very good agreement with experimental ones, thus confirming that there are no visible effects of the alignment in muon groups for all threshold energies. In fact, aligned events do exist as the curve tails in Fig.1 demonstrate it at least for small multiplicity *m*=4 (1 percent of the events with λ=0.9 ÷1.0), but, very similar spots configurations can be realised by chance. Histogram plots obtained with pure random track spread confirm it clearly.

We also used one more way of finding excess of aligned events. Distributions on λ were obtained using the same data of multiplicity **m** by rejecting one trajectory to maximise λ. In other words, we selected groups of **m-1** muons with maximal λ among **m** muons trying to search for aligned configuration of smaller multiplicity. An example of such procedure result is shown in Fig.1d for **m**=5 and for $E_{th}$=0.85. Again a comparison with distributions of random spots (histograms) that had undergone the same procedure didn't give an excess of aligned events either.

## 4. Simulations

A Monte Carlo simulation of EAS propagation in the atmosphere has been carried out. Muon groups were first simulated in the framework of the MC0 code [6] (close to QGSJET), which does not include unusual processes and reproduces well results obtained by the PAMIR Collaboration in emulsion chamber experiments and related to energies $E_0 < 5·10^{15}$ eV. To analyze the alignment phenomenon, a simplified model of coplanar particle generation (CPGM) [7] in the first interactions of primary protons with air nuclei at $E_0 > E_{0\,copl} = 8·10^{15}$ eV has been used. This model characterized with a mean transverse momentum of the most energetic particles of 2.34 GeV/c transversely to the coplanarity plane, reproduces PAMIR's data on aligned γ-h families. The dependence on the detector's geometrical size was considered. It was shown that at $E_{thr} \cong 1$ TeV a fraction of aligned muon groups F($\lambda_{N\mu} \geq 0.6$) simulated in the framework of both the MC0 and CPGM algorithms coincide for all the muon multiplicity, while there exists a dependence on the detector dimensions. For ~10-m detector (BUST dimensions) it is close to that being characteristic for γ-h families. As one can see, a fraction of aligned events on a level of about few percents is normal and this is in agreement with our experimental data. This result could be explained as follows. In accordance with the PAMIR's concept assuming a coplanar generation of several most energetic particles, which cannot actually decay into muons, the appearance of coplanar production does not affect on features of muon groups formed mainly by muons produced lower by lower-energy secondary particles. In this work we analyzed the dependence of results on the muon energy thresholds. Higher threshold corresponds to the higher mean effective energy of primary particle $\langle E_{0\,eff}\rangle$ responsible for production of the recorded muon group.

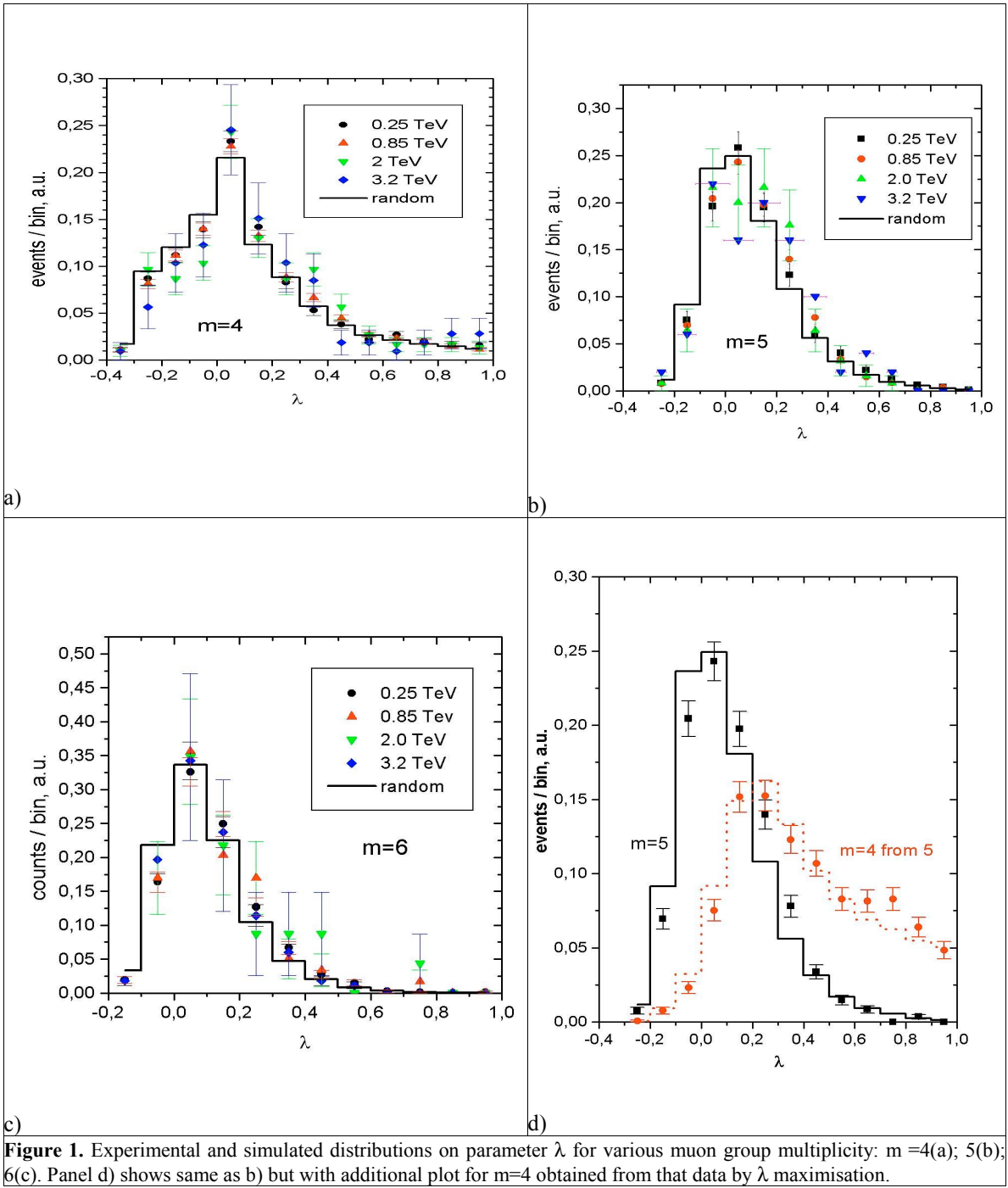

**Figure 1.** Experimental and simulated distributions on parameter λ for various muon group multiplicity: m =4(a); 5(b); 6(c). Panel d) shows same as b) but with additional plot for m=4 obtained from that data by λ maximisation.

Figure 2 shows distributions on λ parameter at $E_{thr}$ = 0.25 ($\langle E_{0\,eff}\rangle < E_{0\,copl}$) and 3.2 ($\langle E_{0\,eff}\rangle > E_{0\,copl}$) TeV for 10-m setups found using MC0 and CPGM. One can see that the form of the distributions is model-independent and depends on energy threshold. In general, the agreement between experimental and simulated data is well. The stronger threshold dependence of simulated distributions shown in Fig. 2 as compared with experimental ones is a result of not taking methodical details into account, which can only reduce all the effects related to the coplanar particle production. Thus, a significant alignment of muons could only be produced in a case of a direct coplanar muon production.

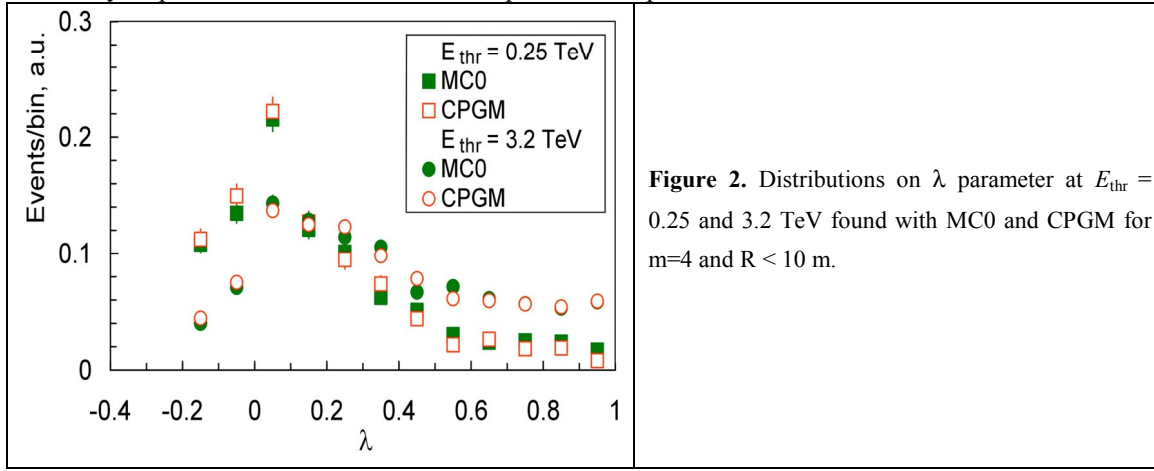

**Figure 2.** Distributions on λ parameter at $E_{thr}$ = 0.25 and 3.2 TeV found with MC0 and CPGM for m=4 and R < 10 m.

## 5. Summary

An experimental search has been performed for aligned events in muon groups with threshold energies equal to 0.25; 0.85; 2.0; 3.2TeV. The measured distributions on the alignment parameter λ agree very well with simulated event distributions and there is no any evidences for the existence of such events among the muon groups underground at a level above the expectations obtained for simulated events with a random track distribution. Moreover, even in EAS's simulated in the framework of a model including the coplanar particle generation there exists no visible effect of alignment in muon groups because only a small fraction of muons could be generated in the first interaction via decay of the most energetic aligned secondaries. Fluctuations of cascade development and a huge number of muons produced in later generations make this process random. Therefore, the observed muon alignment is only determined by random fluctuations of muon tracks on the detector area. The biggest effect of muon alignment could be expected for the highest muon threshold energy and for the lowest **m**. Taking into account statistics, the detector area and a duration of the experiment we can put an upper limit to the flux of aligned muon groups (for maximal threshold energy equal to 3.2 TeV and for **m**=4) on a one sigma level as $F_{aligned} < 2.3 \cdot 10^{-14}$ cm$^{-2}$ sec$^{-1}$ ster$^{-1}$.

\*\*\*

This work was supported in part by the Russian Foundations for Basic Research, projects No 04-02-17083a, 03-02-16272a, 03-02-17465a, 05-02-16781a and Ministry of Education and Science project SSL-1782.2003.2.